\documentclass[onecollarge,natbib]{svjour2}
\bibpunct{[}{]}{;}{n}{}{,} 
\smartqed  
\usepackage{graphicx}
\usepackage{braket}

\renewcommand{\d}{\ensuremath{\mathrm{d}}}

\journalname{Few-Body Systems}
\begin{document}

\title{Gluons at finite temperature
\thanks{Talk presented by P. J. Silva at the Light Cone 2016 Conference,
Lisbon, Portugal, 5-8 September 2016. }
}


\author{P. J. Silva \and O. Oliveira \and D. Dudal 
\and P. Bicudo \and N. Cardoso 
}


\institute{P. J. Silva, O. Oliveira \at CFisUC, Department of Physics, University of Coimbra, P-3004-516 Coimbra, Portugal \\
\email{orlando@fis.uc.pt, psilva@uc.pt}      
\and
D. Dudal \at
KU Leuven Campus Kortrijk - KULAK, Department of Physics, Etienne Sabbelaan 51 bus 7800, 8500 Kortrijk, Belgium and \at Ghent University, Department of Physics and Astronomy, Krijgslaan 281-S9, 9000 Gent, Belgium    \\
\email{david.dudal@kuleuven.be}      
\and
P. Bicudo, N. Cardoso \at
CFTP, Departamento de F\'isica, Instituto Superior T\'ecnico, Universidade de Lisboa, Avenida Rovisco Pais 1, 1049-001 Lisboa, Portugal \\
\email{bicudo@tecnico.ulisboa.pt, nunocardoso@cftp.ist.utl.pt}
}

\date{Received: date / Accepted: date}

\maketitle

\begin{abstract}
The gluon propagator is investigated at finite temperature via lattice simulations. In particular, we discuss its interpretation as a massive-type bosonic propagator. Moreover, we compute the corresponding spectral density and study the violation of spectral positivity. Finally, we explore the dependence of the gluon propagator on the phase of the Polyakov loop.
\end{abstract}

\section{Introduction}
\label{intro}
The study of the dynamics of QCD at finite temperature and density has been the subject of an intensive study, motivated by the heavy-ion experiments running e.g. at CERN \cite{cernexp} and RHIC \cite{rhicexp}. From the theoretical point of view, the lattice formulation has been one of the most promising frameworks which allows to investigate the properties of the non-perturbative regime of QCD at non-zero temperature.
For pure SU(3) Yang-Mills theory at finite temperature, a first-order transition is found at a critical temperature $T_c\sim 270$ MeV --- see, for example, \cite{Tc} and references therein. For temperatures below $T_c$, the gluons are confined within color-singlets, whereas for $T>T_c$ the gluons become deconfined and behave as massive quasiparticles. 
The order parameter for the deconfinement phase transition is the Polyakov loop, defined on the lattice as
\begin{equation}
L( \vec{x} ) = \mathrm{Tr} \prod^{N_t-1}_{t=0} \, \mathcal{U}_4(\vec{x},t) \quad , \quad L = \langle L( \vec{x} ) \rangle \, \propto \, e^{-F_q/T}
\end{equation}
where $\mathcal{U}_4$ is the time-oriented link. Its space-averaged value $L$
is a measure of the free energy of a static quark. 
The behaviour of the Polyakov loop as a function of the temperature is connected with a spontaneous breaking of the center symmetry. The Wilson gauge action and the integration measure are invariant under a center transformation, where the temporal links on a hyperplane $x_4=const$ are multiplied by an element of the SU(3) center group $z \in Z_3 = \{e^{- i 2 \pi/3}, 1, e^{ i 2 \pi/3} \}$. Under such a center transformation, the Polyakov loop transforms as $L(\vec{x}) \rightarrow z L(\vec{x})$. For temperatures below $T_c$, the local phase of the Polyakov loop is equally distributed among the three sectors, and therefore  $L = \langle L( \vec{x} ) \rangle \approx 0$. For $T>T_c$, the $Z_3$ sectors are not equally populated (a manifestation of a spontaneous breaking of the center symmetry) and thus $L \neq 0$.

In this proceeding, we focus on several aspects of the Landau gauge gluon propagator at finite temperature, computed via lattice QCD simulations. Like other propagators of fundamental fields (e.g. quark and ghost propagators), gluon two-point functions encode information about non-perturbative phenomena, such as confinement and deconfinement. 

\section{Lattice setup and propagators}

At finite temperature, the gluon propagator has the following tensor structure 
\begin{equation}
D^{ab}_{\mu\nu}(\hat{q})=\delta^{ab}\left(P^{T}_{\mu\nu} D_{T}(q_4,\vec{q})+P^{L}_{\mu\nu} D_{L}(q_4,\vec{q}) \right) 
\label{tens-struct}
\end{equation}
where $P^{T}$ and $P^{L}$ are the transverse and longitudinal projectors respectively:
\begin{equation}
P^{T}_{\mu\nu} = (1-\delta_{\mu 4})(1-\delta_{\nu 4})\left(\delta_{\mu \nu}-\frac{q_\mu q_\nu}{\vec{q}^2}\right) \quad , \quad P^{L}_{\mu\nu} = \left(\delta_{\mu \nu}-\frac{q_\mu q_\nu}{{q}^2}\right) - P^{T}_{\mu\nu} \, .
\label{gluon-proj}
\end{equation}

\begin{table}[t]
\caption{Lattice setup.}
\centering
\label{lattsetup}
\begin{tabular}{cccccc}
\hline\noalign{\smallskip}
Temp. (MeV) &    $\beta$ & $L_s$ &  $L_t$ & a [fm] & 1/a (GeV) \\[3pt]
\tableheadseprule\noalign{\smallskip}
121 &    6.0000 & 64    &     16 &     0.1016 &     1.943 \\
162 &    6.0000 & 64    &     12 &     0.1016 &     1.943 \\
194 &    6.0000 & 64    &     10 &     0.1016 &     1.943 \\
243 &    6.0000 & 64    &     8 &     0.1016 &     1.943 \\
260 &    6.0347 & 68    &     8 &     0.09502 &     2.0767 \\
265 &    5.8876 & 52    &     6 &     0.1243 &     1.5881 \\
275 &    6.0684 & 72    &     8 &     0.08974 &     2.1989 \\
285 &    5.9266 & 56    &     6 &     0.1154 &     1.7103 \\
290 &    6.1009 & 76    &     8 &     0.08502 &     2.3211 \\
305 &    5.9640 & 60    &     6 &     0.1077  &      1.8324 \\
305 &    6.1326 & 80    &     8 &     0.08077 &     2.4432 \\
324 &    6.0000 & 64    &     6 &     0.1016  &      1.943 \\
366 &    6.0684 & 72    &     6 &     0.08974 &     2.1989 \\
397 &    5.8876 & 52    &     4 &     0.1243 &     1.5881 \\
428 &    5.9266 & 56    &     4 &     0.1154 &     1.7103 \\
458 &    5.9640 & 60    &     4 &     0.1077  &      1.8324 \\
486 &    6.0000 & 64    &     4 &     0.1016  &      1.943 \\
\hline\noalign{\smallskip}
\end{tabular}
\end{table}

In Table \ref{lattsetup} we describe our lattice setup used in the next two sections. The temperature is defined by adjusting the lattice temporal size, $T=1/(N_ta)$. Following our first reports \cite{rep2012}, where we saw a measurable dependence of the gluon propagator at finite $T$ on the lattice volume, we have carefully chosen the lattice parameters of the various Monte Carlo simulations in order to keep the physical (spatial) volume at a constant value $\sim (6.5\,\mathrm{fm})^3$. Simulations have been made in Coimbra \cite{lca} with the help of Chroma \cite{chroma} and PFFT \cite{pfft} libraries.
 
The results shown here are for renormalized longitudinal and transverse propagators, and for a renormalization scale $\mu=4\,$GeV. For details about the renormalization procedure see \cite{gluonmass}.

\begin{figure*}
\centering
  \includegraphics[width=0.45\textwidth]{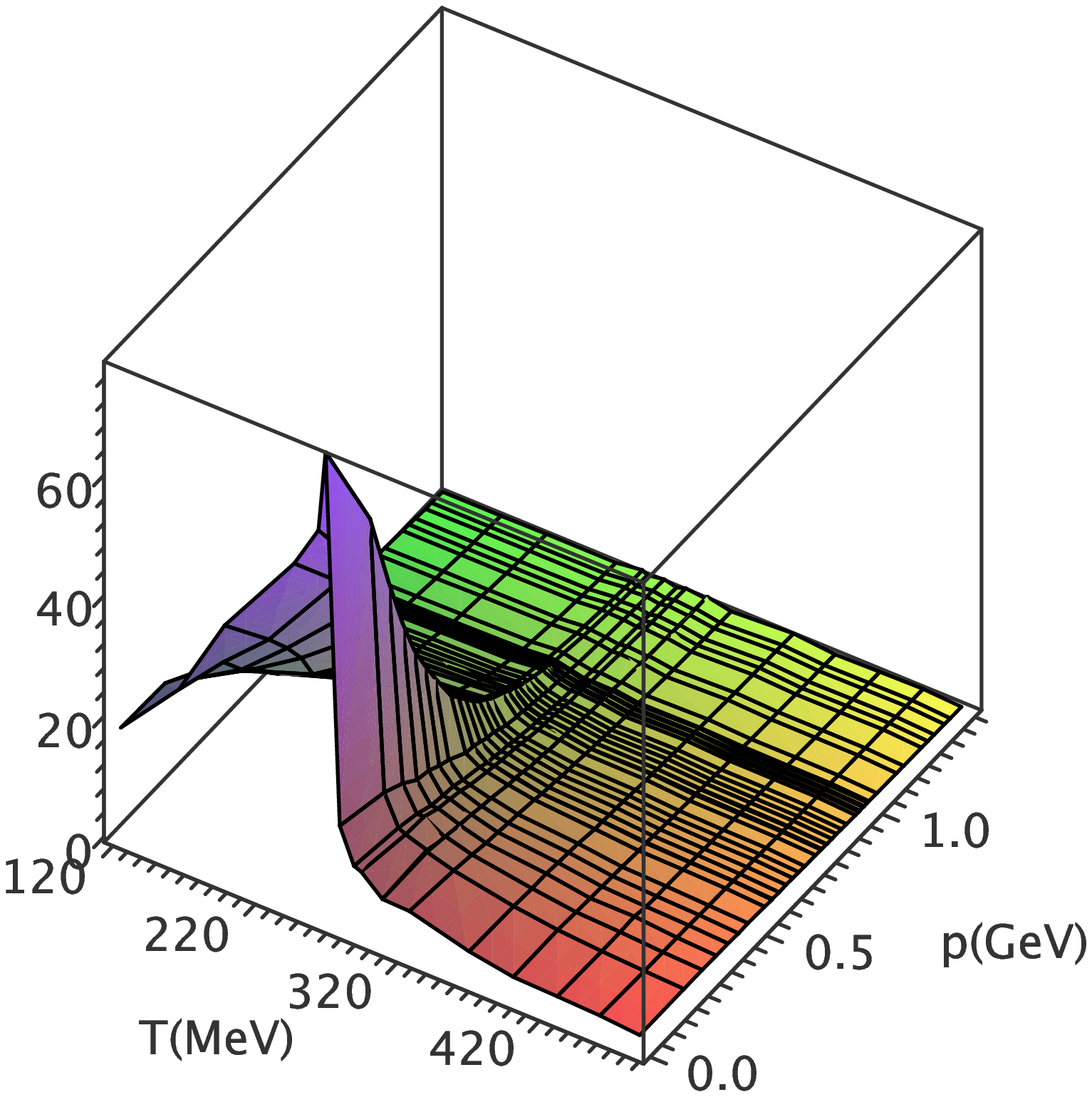} \quad
  \includegraphics[width=0.45\textwidth]{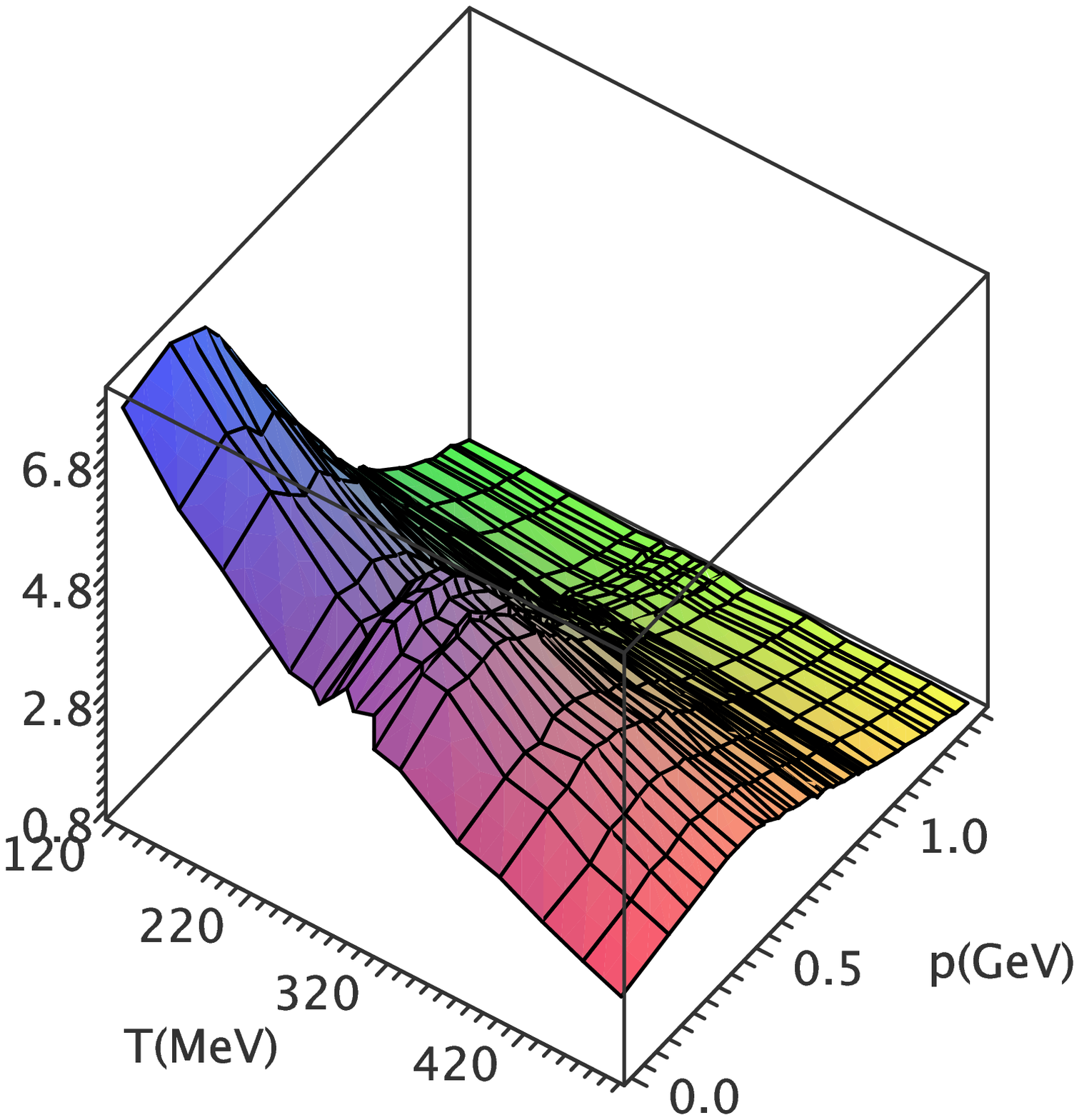}
\caption{Longitudinal (left) and transverse (right) components of the gluon propagator as functions of momentum and temperature.}
\label{plot3d}       
\end{figure*}

In Fig. \ref{plot3d} we show how the propagators behave as functions of momentum and temperature. Note the sharp transition for the longitudinal component at $T\sim T_c$, and the turnover of the tranverse component in the infrared region, for $T \gg T_c$.

%

\section{Positivity violation and spectral densities}

It is well known that a Euclidean momentum-space propagator of a (scalar) physical degree of freedom 
\begin{equation}
\mathcal{G}(p^2)\equiv\braket{\mathcal{O}(p)\mathcal{O}(-p)}
\end{equation}
oughts to have a K\"{a}ll\'{e}n-Lehmann spectral representation 
\begin{equation}
\mathcal{G}(p^2)=\int_{0}^{\infty}\d\mu\frac{\rho(\mu)}{p^2+\mu}
\,,\qquad \textrm{with }\rho(\mu)\geq0 \textrm{ for } \mu\geq 0\,.
\end{equation}
where the spectral density $\rho(\mu)$ contains information on the masses of physical states described by the operator $\mathcal{O}$. In \cite{spectral} a method is presented which allows to compute the spectral density of gluons and other (un)physical degrees of freedom, for which the spectral density is not strictly positive. The method relies on Tikhonov regularization combined with the Morozov discrepancy principle. Here we discuss some preliminary results \cite{latt2013} for the spectral density associated to the gluon propagator at finite temperature, together with the temporal correlator
\begin{equation}
  C(t) = \int_{-\infty}^{\infty} \frac{dp}{2\pi} D(p^2) \exp(-ipt)=  \int_{0}^{\infty} d\omega \rho(\omega^2) e^{-\omega t}.
\end{equation}
Note that $C(t) < 0$ in some range of $t$ implies a negative spectral density, hence positivity violation and gluon confinement. On the other hand, a positive $C(t)$ says nothing about the sign of $\rho(\mu)$.

\begin{figure*}[t]
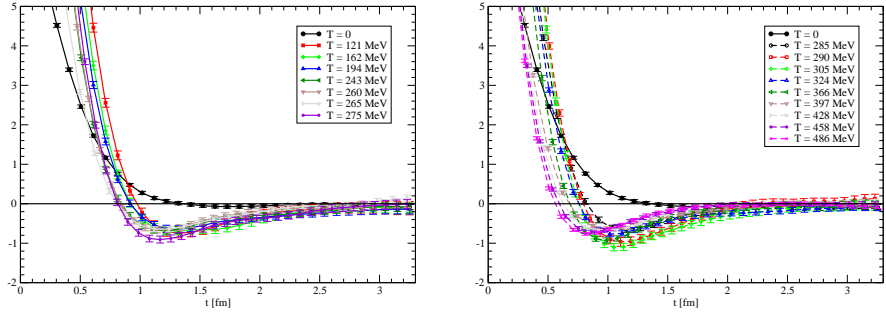

\vspace*{0.5cm}
\centering
  \includegraphics[width=0.35\textwidth]{positividade_trans_upTc} \quad\quad
  \includegraphics[width=0.35\textwidth]{positividade_trans_aboveTc}
\vspace*{0.5cm}
\caption{Temporal correlator for the transverse component.}
\label{postrans}       
\end{figure*}

\begin{figure*}[t]
\vspace*{-0.4cm}
\centering
  \includegraphics[width=0.75\textwidth]{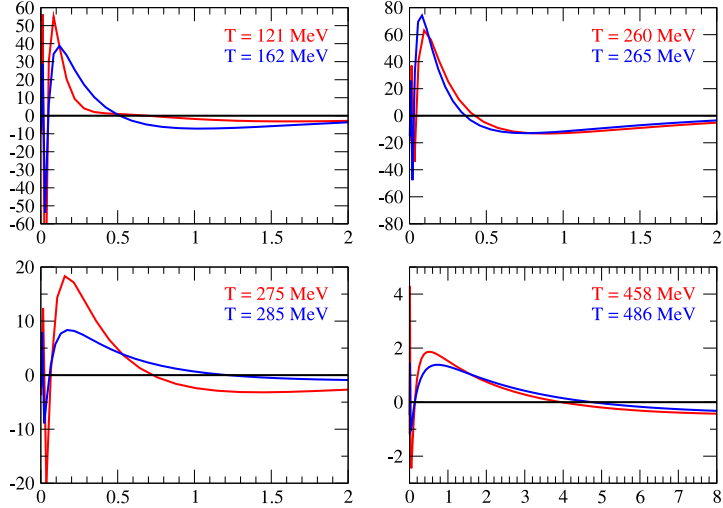}
\caption{Spectral densities for the longitudinal component.}
\label{speclong}       
\end{figure*}

In Fig. \ref{postrans} we plot $C(t)$ for the tranverse component. We conclude that the positivity is violated for all temperatures. Furthermore, a careful inspection reveals that the time scale for positivity violation decreases with the temperature. In Fig. \ref{speclong} we plot, for a number of selected temperatures, the spectral density of the longitudinal component. The plots show that the momentum scale at which the spectral density becomes negative seems to increase with the temperature. All these results suggests that, for sufficiently high temperatures, the spectral density may be strictly positive and, therefore, gluons would behave as quasi-particles.

%
%

\section{Gluon mass}

In this section we investigate whether the gluon propagator at finite temperature behaves as a massive-type bosonic propagator \cite{gluonmass}. We consider a Yukawa-type ansatz
\begin{equation}
 D(p) = \frac{Z }{ p^2 + M^2} 
\label{yukawa}
\end{equation}
where $M$ is the gluon mass and $Z^{\frac{1}{2}}$ the overlap between the gluon state and the quasi-particle massive state.

\begin{figure*}[t]
\centering
  \includegraphics[width=0.45\textwidth]{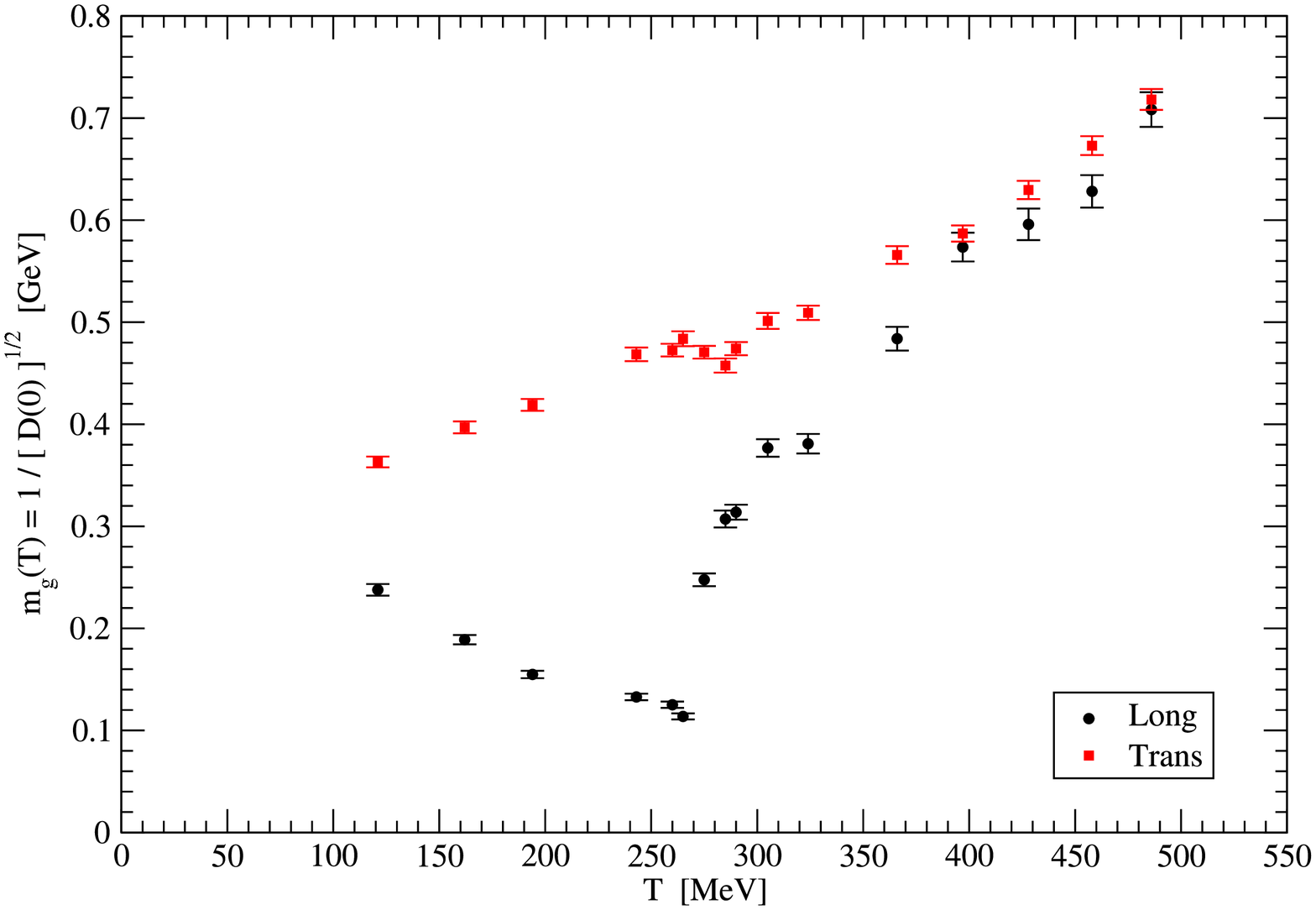} \quad\quad
  \includegraphics[width=0.45\textwidth]{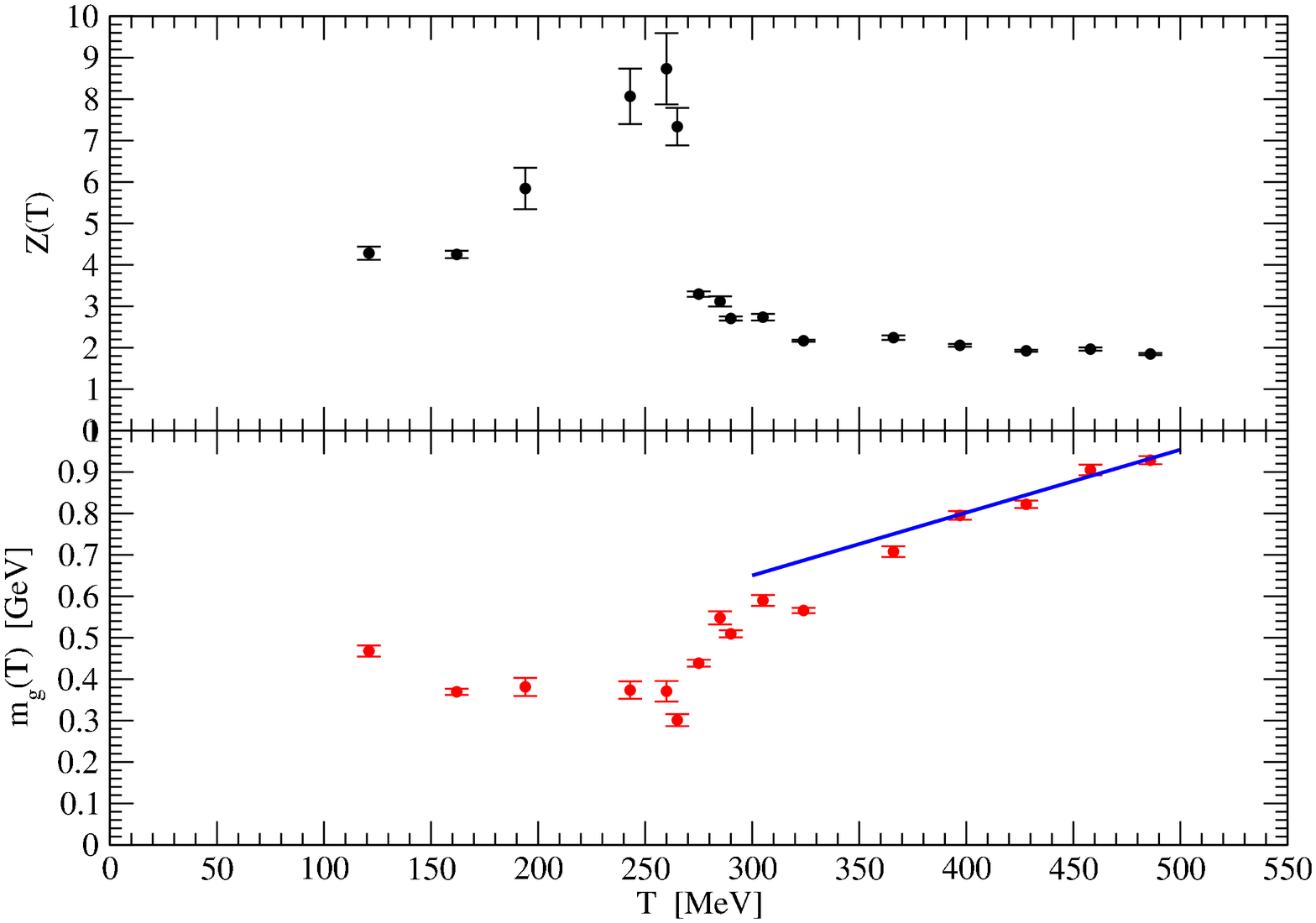}
\caption{Gluon mass scales as function of the temperature.}
\label{gmass}       
\end{figure*}

The simplest definition for a gluon mass scale is given by $M = 1 / \sqrt{ D(0)} $. Our results for a mass scale considering this definition can be seen in the left plot of Fig. \ref{gmass}. Of course, a more realistic estimate for the gluon mass can be obtained by fitting our lattice data in the infrared region to the ansatz described in Eq. (\ref{yukawa}). It turned out that the lattice data for the transverse propagator is not compatible with such ansatz. The results for the longitudinal propagator can be seen in the right plot of Fig. \ref{gmass}. 

%

\section{$Z_3$ dependence}

As seen in Fig. \ref{plot3d}, $D_L$ and $D_T$ show quite different behaviours with $T$. The gluon propagator is usually computed such that $-\pi/3<\arg(L)\leq\pi/3$ i.e. in the so-called $Z_3$ sector 0. In this section we investigate the behaviour of the gluon propagator in the  other $\pm1$ $Z_3$ sectors\footnote{The $Z_3$ sector $-1$ corresponds to $-\pi<\arg(L)\leq-\pi/3$, and the $Z_3$ sector $1$ is defined by $\pi/3<\arg(L)\leq\pi$.} and compare with the results for the zero sector. To achieve such goal, for each configuration in a given ensemble, we applied a center transformation considering all $z\in Z_3$, thus obtaining 3 different configurations related by $Z_3$ transformations, with the very same value for the Wilson action. Each configuration is then rotated to the Landau gauge. Finally, each of the three gauge configurations is classified according to the phase of $L=|L|e^{i\theta}$.  

\begin{figure*}[t]
\vspace*{0.5cm}
\centering
  \includegraphics[width=0.35\textwidth,angle=-90]{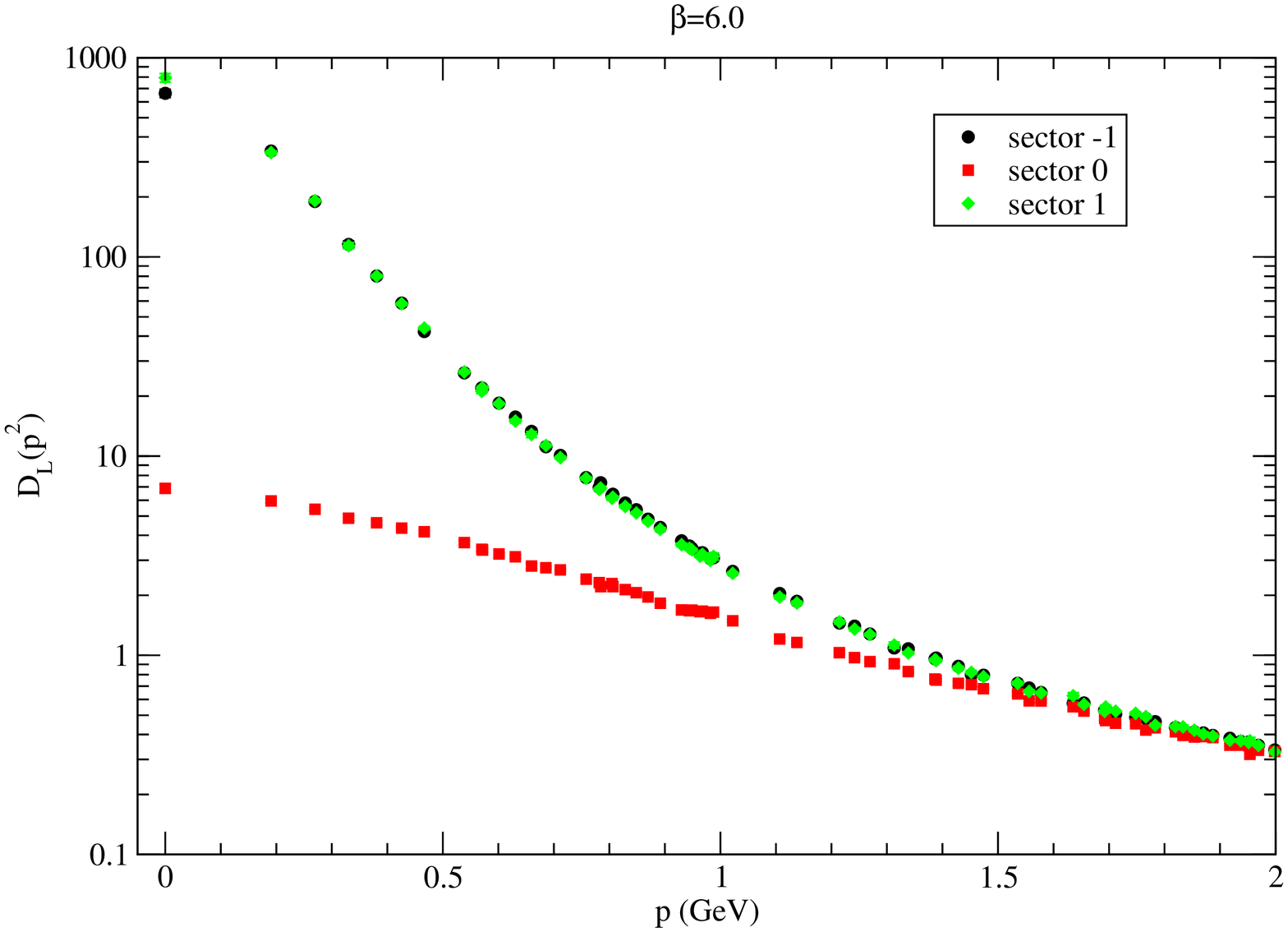} \quad\quad
  \includegraphics[width=0.35\textwidth,angle=-90]{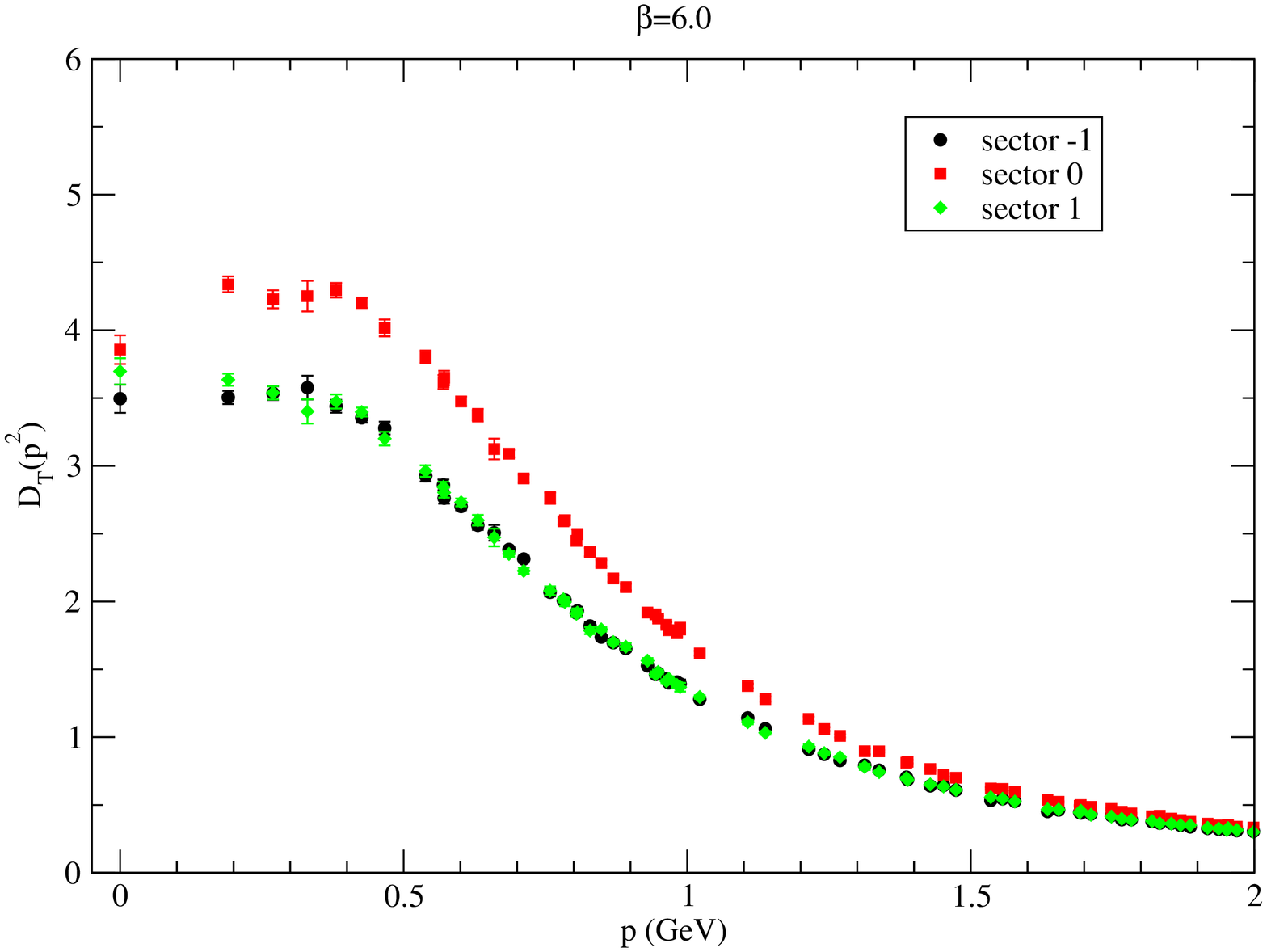}
\vspace*{0.5cm}
\caption{Longitudinal (left) and transverse (right) propagators for the different sectors at $T=324\,$MeV.}
\label{z3-324}       
\end{figure*}

\begin{figure*}[t]
\vspace*{0.5cm}
\centering
  \includegraphics[width=0.35\textwidth,angle=-90]{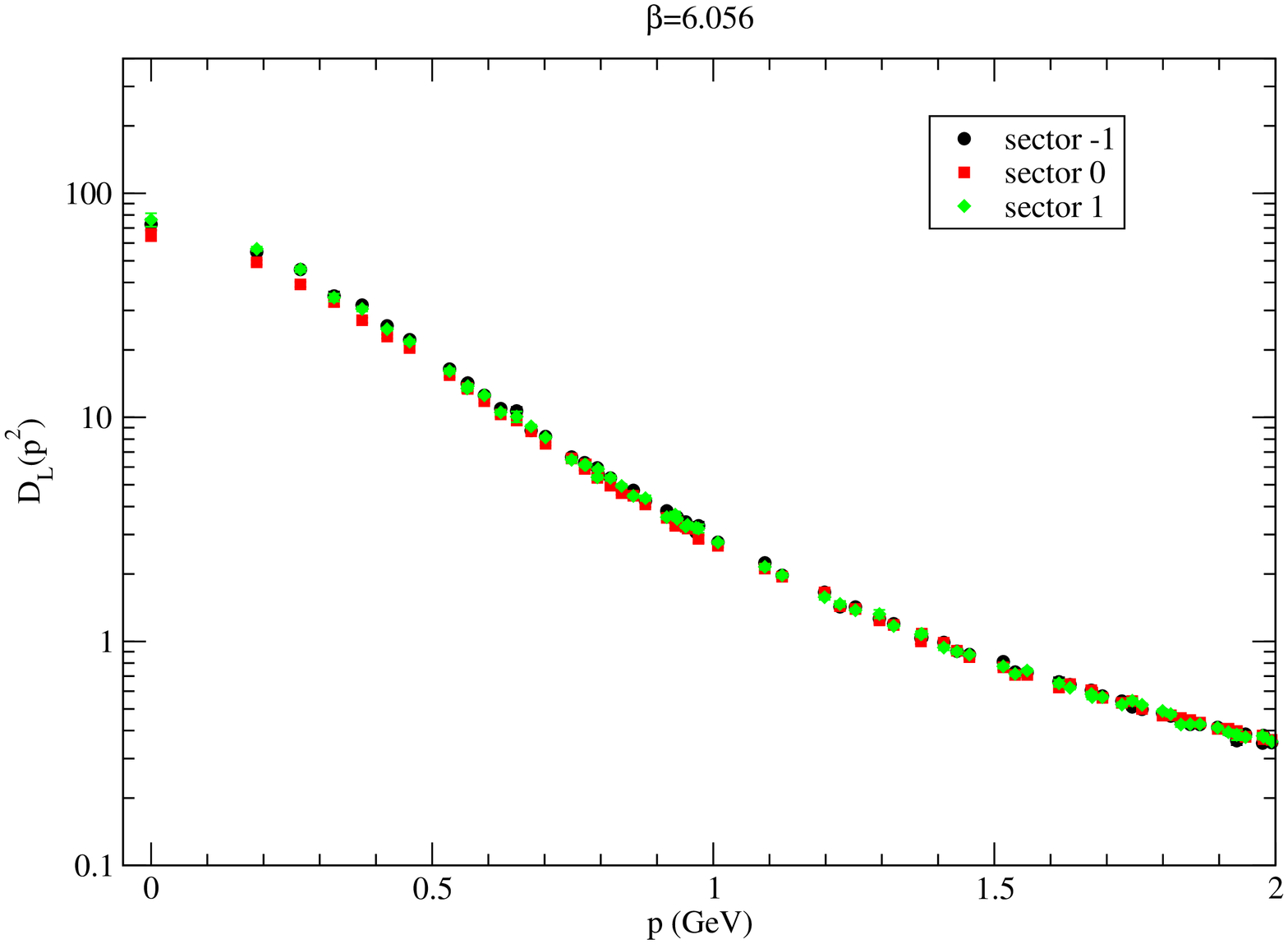} \quad\quad
  \includegraphics[width=0.35\textwidth,angle=-90]{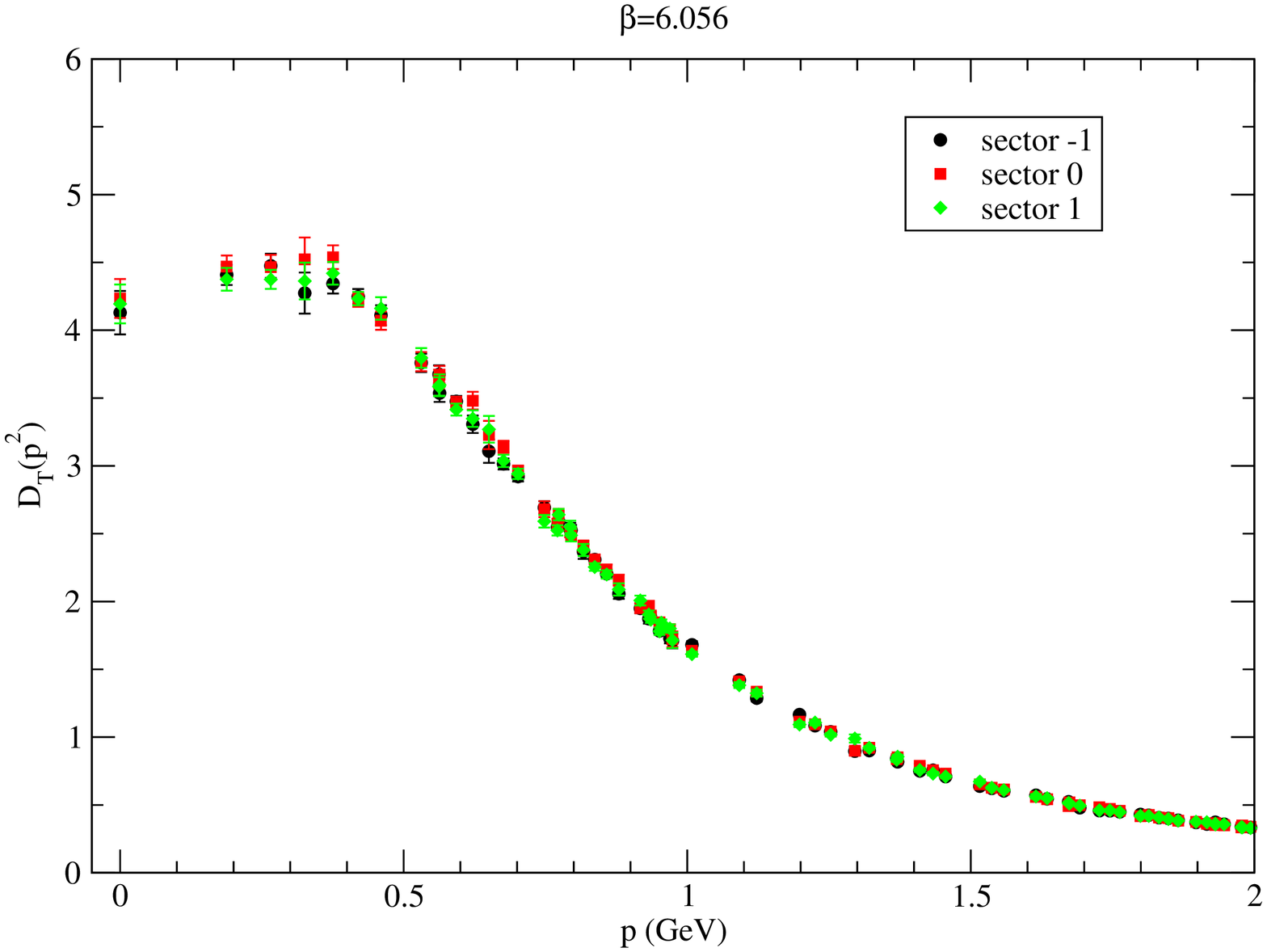}
\vspace*{0.5cm}
\caption{Longitudinal (left) and transverse (right) propagators for the different sectors for $T=269\,$MeV, slightly below $T_c$.}
\label{z3-269}       
\end{figure*}

In Fig. \ref{z3-324}, we can see the typical behaviour of the gluon propagator in the different $Z_3$ sectors for a temperature well above $T_c$. For the longitudinal component, the propagator in the $\pm1$ sectors is strongly enhanced relative to the $0$ sector. On the other hand, the tranverse propagator in the $\pm1$ sectors is suppressed if compared with the $0$ sector. However, for temperatures below $T_c$, the picture changes --- see Fig.  \ref{z3-269}. In this case, the three propagators for the different $Z_3$ sectors are indistinguishable. 

\begin{figure*}[t]
\vspace*{0.5cm}
\centering
  \includegraphics[width=0.35\textwidth,angle=-90]{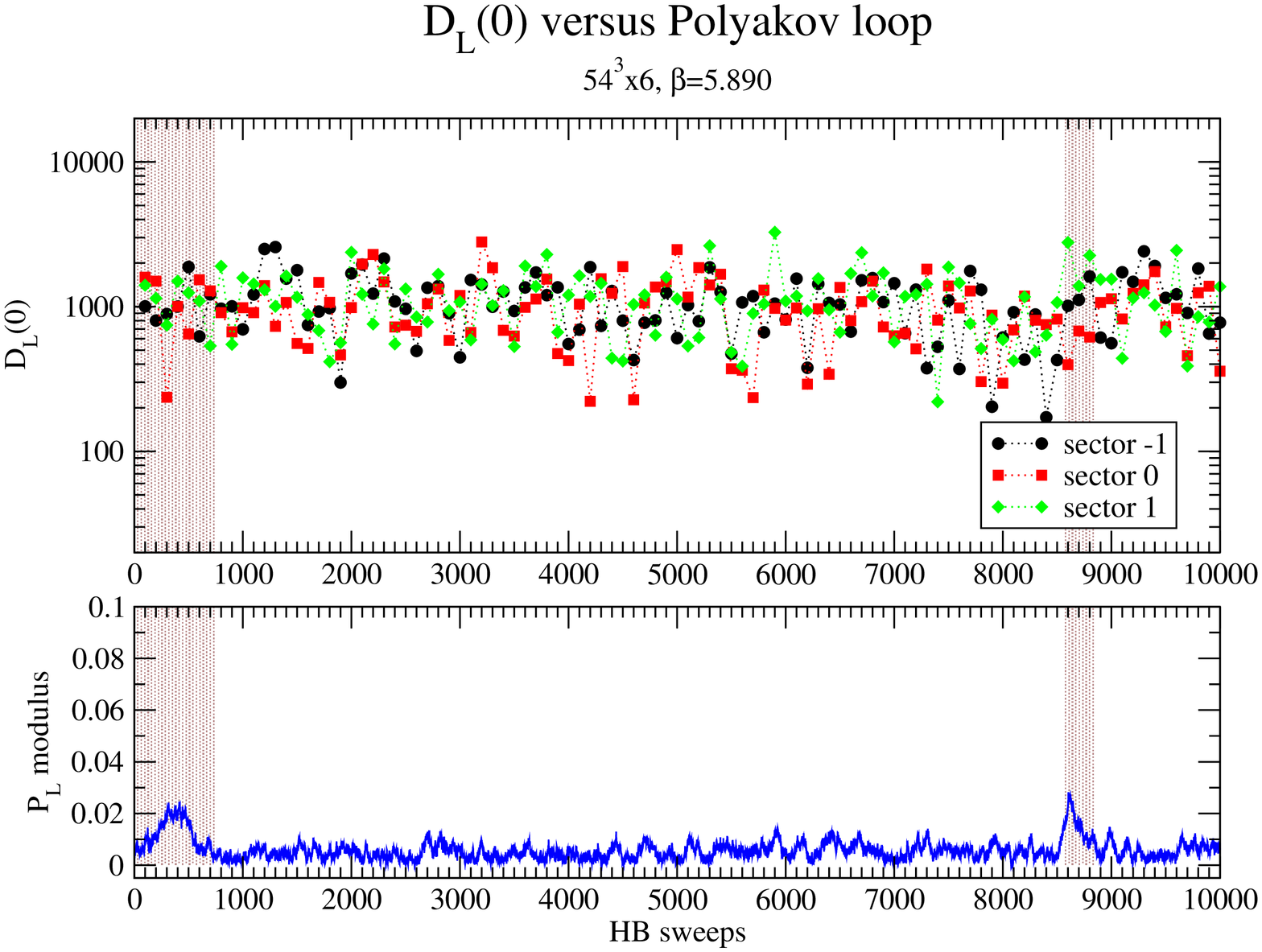} \quad\quad
  \includegraphics[width=0.35\textwidth,angle=-90]{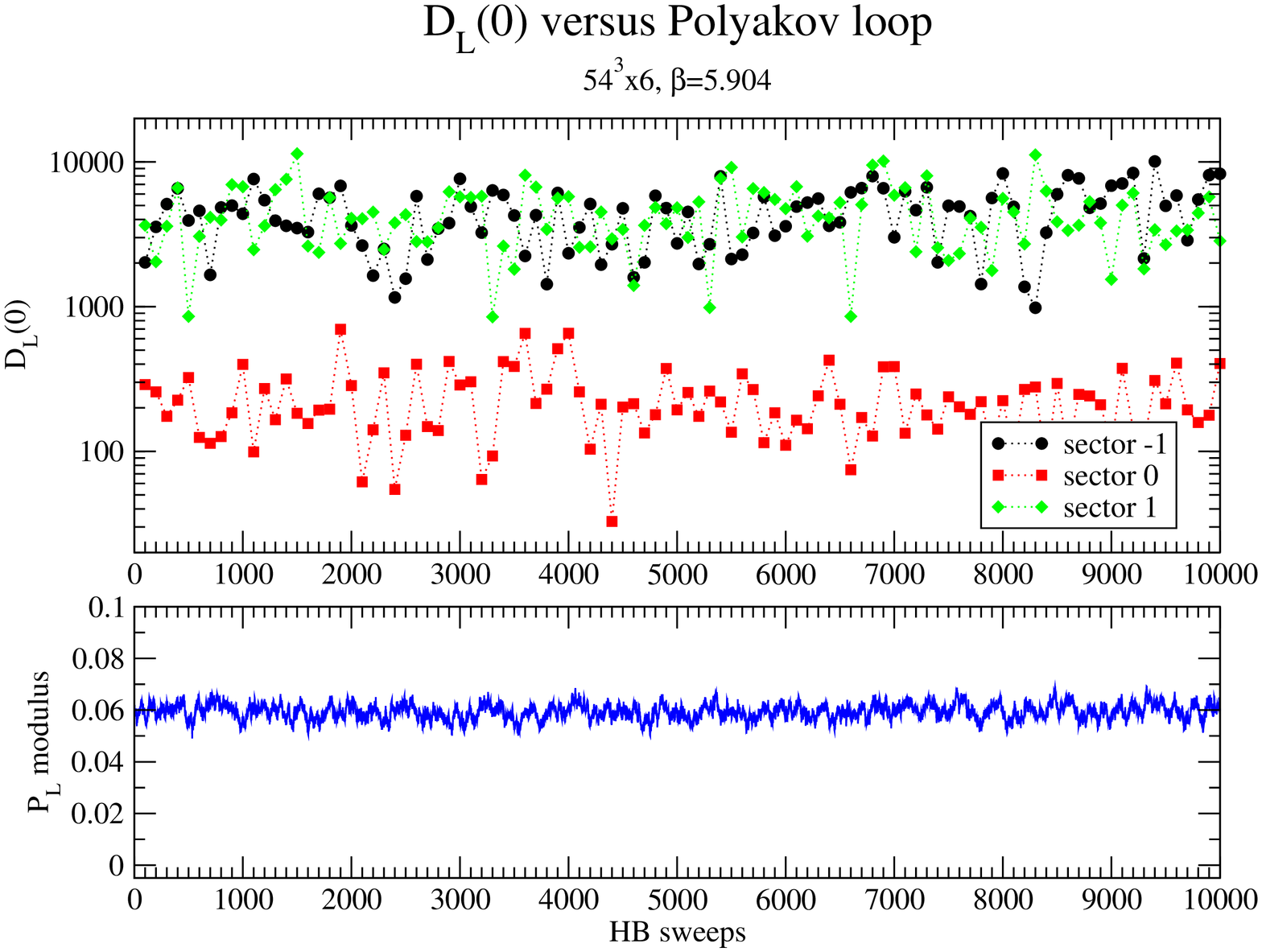}
\vspace*{0.5cm}
\caption{Monte Carlo history of the bare values of $D_L(0)$ and $|L|$ for ensembles with $T= 266\,\mathrm{MeV}<T_c$ (left)  and $T= 273\,  \mathrm{MeV}>T_c$ (right). }
\label{z3-mc}       
\end{figure*}

A comparison of the Markov chain history for temperatures below and above $T_c$ --- see Fig. \ref{z3-mc} --- allows one to conclude that the difference of the longitudinal propagator between the different $Z_3$ sectors can be used as a criterion to identify whether a given configuration is in the confined or deconfined phase. 

More details about this work may be found in \cite{z3}.

%
%

%
%

%

\begin{acknowledgements}
O. Oliveira, and P. J. Silva acknowledge financial support from FCT Portugal under contract with reference UID/FIS/04564/2016. P. J. Silva acknowledges support by FCT under contracts SFRH/BPD/40998/2007 and SFRH/BPD/109971/2015. P. Bicudo and N. Cardoso thank CFTP with FCT grant UID/FIS/00777/2013. N. Cardoso is supported by FCT under contract SFRH/BPD/109443/2015. The computing time was provided by the Laboratory for Advanced Computing at the University of Coimbra \cite{lca}.
\end{acknowledgements}

\newpage

\end{document}